\documentclass[aps,11pt,pra,showpacs,superscriptaddress,preprint]{revtex4}

\usepackage{amsmath}

\usepackage[dvipdfmx]{graphicx}

\usepackage{subfig}

\usepackage{array}

\usepackage{amsfonts}

\usepackage{epsf}

\usepackage{epsfig}

\usepackage{amssymb}

\usepackage{bbm}

\usepackage{delarray}

\usepackage{caption}

\usepackage{amstext}

\usepackage{graphics}

\usepackage{hyperref}
\hypersetup{
colorlinks=true,
urlcolor=blue,
citecolor=blue}
\usepackage{natbib}

\catcode`ð=\active

\defð{\u{g}}

\catcode`Ð=\active

\defÐ{\u{G}}

\catcode`Ý=\active

\defÝ{\. I}

\catcode`ö=\active

\defö{\"{o}}

\catcode`Ö=\active

\defÖ{\"O}

\catcode`ü=\active

\defü{\"{u}}

\catcode`Ü=\active

\defÜ{\"{U}}

\catcode`Þ=\active

\defÞ{\c{S}}

\catcode`þ=\active

\defþ{\c{s}}

\catcode`ý=\active

\defý{{\i}}

\catcode`ç=\active

\defç{\d{c}}

\catcode`Ç=\active

\defÇ{\d{C}}

\begin{document}

\title{Three-Dimensional Eigenvalues of Harmonic Oscillator- and Coulomb-type Potentials from One-Dimensional Generalized Morse Potential: Perturbative Analyse based on Generalized Laguerre Polynomials}

\author{\small Altuð Arda}
\email[E-mail: ]{arda@hacettepe.edu.tr}\affiliation{Division of
Physics Education, Hacettepe University, 06800, Ankara,Türkiye}

\begin{abstract}
We present perturbative energy eigenvalues (up to second order) of Coulomb- and harmonic oscillator-type fields within a perturbation scheme. We have the required unperturbed eigenvalues ($E_{n}^{(0)}$) analytically obtained by using similarities between the expressions obtained from unperturbed Hamiltonian(s) for two fields and obtained from the ones for one-dimensional generalized Morse field. We use the Langer transformation for this aim. We need the diagonal and non-diagonal matrix elements of unperturbed and perturbed Hamiltonians to get energy eigenvalues perturbatively, which are obtained with help of some recursion identities or some integrals of generalized Laguerre polynomials having analytical results.
\end{abstract}


\maketitle

\newpage

\section{Introduction}
The integrals having Laguerre polynomials with analytical/numerical results are crucial in physics and chemistry: normalization of wave function \cite{MN, LD, HDM}, the expectation values of position in Coulomb problem where integrals of Laguerre polynomials are written in terms of Hahn and Chebyshev polynomials \cite{ST}, determining the information measures, Fisher's, Renny's and Shannon's entropies \cite{YAD}, spready measures\cite{SMD, TY}, probability distributions \cite{DLOY} within quantum information theory. They are also related with QED corrections to bound state energies for Coulomb field \cite{AALS}. The Laguerre polynomials and analytical results of their integrals are also a mathematical goal in literature \cite{MA, SMA}. In Ref \cite{AA}, perturbative corrections of a neutral, relativistic particle subjected to electromagnetic field are studied, where analytical results of some integrals including Laguerre polynomials or their derivatives are given as a list.

The solutions, which are explicit and analytic, of dynamical systems in the quantum theory are possible only for a few system whose potential fields may be harmonic oscillator, constant, linear and Coulomb potentials \cite{PAMD, M}. Most of the quantum mechanical problems cannot be studied exactly because they are described by equations whose solutions are not expressed in finite terms. One offers then perturbation methods \cite{F1} to solve for such problems \cite{PAMD}. The general framework of them is based on the idea that the total Hamiltonian of the system can be divided into two parts, one of which is called unperturbed part having exact, analytical solutions and the other is small (or perturbed) part of total Hamiltonian, and leads to small corrections in solution \cite{PAMD}. This is also valid for standard perturbation theory which described on textbook level in details \cite{M, TD}, and also used in literature \cite{FS,F2,V,AA}. In the present work, we study the small corrections on energy eigenvalues of the systems described by Coulomb-like and harmonic oscillator-like potentials within perturbation scheme. We need to use some recursion identities or analytic results of some integrals of generalized Laguerre polynomials to obtain perturbative terms coming from first- and second-order corrections. We calculate the required integrals of generalized Laguerre polynomials with the help of generating function $U_i(z,f)$ \cite{TD}. About obtaining analytical solutions a new area appears in last decades where the Heun differential equation is linked to some relativistic and/or non-relativistic quantum mechanical problems \cite{TM, M1}. The solution of some physical problems in terms of the Heun functions has been studied by many groups \cite{ATT,QG,CX,N,SG,XC,LB,CY1,CY2,GCH}.

It is known that the bounded energies of a non-relativistic particle moving in Morse field are studied by many authors \cite{AS, GCAC,NC,NCP}. de Castro and co-workers are presented exact, analytical solutions of non-relativistic generalized Morse potential in one-dimension in \cite{NC}, where they are obtained some similarities between the equation coming from the Hamiltonian for generalized Morse potential and the equations coming from Hamiltonians for Coulomb and harmonic oscillator potentials in three-dimension to obtain the energy eigenvalues and the corresponding wave functions. We obtain unperturbed eigenvalues required for the perturbation scheme of Coulomb and harmonic oscillator-like potentials from eigenvalues of one-dimensional Morse potential by using the Langer transformation \cite{NC}. It makes possible to study the perturbative corrections to energy, where we use analytical results of some integrals involving generalized Laguerre polynomials or some recursion relation(s) of them.

The work is presented as follows. In Section 2 we obtain unperturbed energy eigenvalues of harmonic oscillator- and Coulomb-like fields in three-dimensions. We write first the eigenvalue equation for the one-dimensional generalized Morse field, and then give equations obtained with the help of Langer transformation which applied on equations coming from unperturbed part of Hamiltonians of the above fields. We modify the Hamiltonians by using the parameter $s$, which is called screening, control or adjustment parameter \cite{FS, GFC}. We write also a term, $h'$, in Hamiltonians as an initial perturbing potential \cite{GFC}. In Section 3 we obtain perturbative solutions up to second-order. We give the results coming from first- and second-order corrections analytically by using some recursion relation(s) or some integrals of the Laguerre polynomials. We summarize also some numerical results in Tables I, and II. Our conclusions are given in last Section.

\section{Unperturbative Solutions}
The time-independent Schrödinger equation is an eigenvalue equation for a non-relativistic particle subjected to the potential $V(\vec{r\,})$
\begin{eqnarray}
\label{eq1}
\left[-\,\frac{1}{2}\,\vec{\nabla\,}^2+V(\vec{r\,})\right]\Psi_{n\ell}(\vec{r\,})=E_{n}^{(0)}\Psi_{n\ell}(\vec{r\,})\,,
\end{eqnarray}
where $E_{n}^{(0)}$ are eigenvalues of the equation, $\Psi_{n\ell}$ is corresponding eigenfunction, and the pair of $(n, \ell)$ are understood as quantum numbers. One expects that energy eigenvalues are real (positive or negative) for bound-states of the particle. Writing the eigenfunction as $\Psi_{n\ell}(\vec{r\,})=r^{-1}\psi_{n\ell}(r)Y_{n\ell}(\theta,\vartheta)$ gives the radial part of the whole equation
\begin{eqnarray}
\label{eq2}
\left[-\,\frac{1}{2}\,\frac{d^2}{dr^2}+V(r)+\,\frac{\ell(\ell+1)}{2r^2}\right]\psi_{n\ell}(r)=E_{n}^{(0)}\psi_{n\ell}(r)\,,
\end{eqnarray}
which is valid for spherically symmetric potential $V(r)$ with $\int_{0}^{\infty}dr|\psi(r)|^2=1$ for bound-state solutions.

We use a generic form for the potential \cite{NC}
\begin{eqnarray}
\label{eq3}
V(r)=sr^{\mu}+\,\frac{\nu}{r^2}\,,
\end{eqnarray}
which corresponds to harmonic oscillator- and Coulomb-like potentials in below.

Within the perturbation theory, the total Hamiltonian is routinely divided into two parts
\begin{eqnarray}
\label{eq4}
h(s)=h_0+h_1\,,
\end{eqnarray}
where the extended unperturbed Hamiltonian is
\begin{eqnarray}
\label{eq5}
h_0=\,-\,\frac{1}{2}\,\frac{d^2}{dr^2}+sr^{\mu}+\,\frac{\nu}{r^2}\,,
\end{eqnarray}
and the extended perturbing potential is
\begin{eqnarray}
\label{eq6}
h_1=(Z-s)r^{\mu}+h'\,.
\end{eqnarray}
with the screening parameter $s$ and $h'$ denotes initial perturbing term \cite{GFC}.

Our aim is to study the perturbative effects on the solutions. So, we need first to obtain the eigenvalues of extended unperturbed Hamiltonian ($h_0$). We give them with the help of similarity between eigenvalue equation for unperturbed Hamiltonian and the ones for Hamiltonian written for one-dimensional Morse potential where we use the Langer transformation.

The Schrödinger equation for the one-dimensional generalized Morse potential is
\begin{eqnarray}
\label{eq7}
\left[-\,\frac{1}{2}\,\frac{d^2}{dx^2}+V_{1}e^{-\alpha x}+V_2e^{-2\alpha x}-\varepsilon_n\right]\phi_{n}(x)=0\,\,\,(\alpha>0)\,,
\end{eqnarray}
where $\varepsilon_n$ is energy eigenvalues and normalization of the corresponding eigenfunctions is $\int_{-\infty}^{\infty}dx|\phi_{n}(x)|^2=1$.

Because of behaviour of the wave function in limits, which also requires $V_2>0$, we write $\phi=y^{\sqrt{-2\varepsilon_n\,}/\alpha}e^{-y/2}f(y)$ with a new variable $y=(2\sqrt{2V_2\,}/\alpha)e^{-\alpha x}$ we obtain a Kummer's-type equation
\begin{eqnarray}
\label{eq8}
\left[y\,\frac{d^2}{dy^2}+\left(\frac{2\sqrt{-2\varepsilon_n\,}}{\alpha}+1-y\right)\frac{d}{dy}-\left(\frac{V_1}{2\sqrt{V_2\,}\alpha}+
\frac{\sqrt{-2\varepsilon_n\,}}{\alpha}+\frac{1}{2}\right)\right]f(y)=0\,,
\end{eqnarray}
whose polynomial solutions are the generalized Laguerre polynomials $L_n^{(k)}(z)$. The complete solution is
\begin{eqnarray}
\label{eq9}
\phi_n(y)=A_n y^{\sqrt{-2\varepsilon_n\,}/\alpha}e^{-y/2}L_n^{(2\sqrt{-2\varepsilon_n\,}/\alpha)}(y)\,,
\end{eqnarray}
with normalization constant $A_n$. The potential parameter $V_1$ should be negative to have finite solutions \cite{NC}. The polynomial solution requires a restriction on the number of allowed states as $n<\frac{|V_1|}{\sqrt{2V_2\,}\alpha}-\frac{1}{2}$ and gives the existed eigenvalues for Eq. (\ref{eq7}) \cite{NC}
\begin{eqnarray}
\label{eq10}
\varepsilon_n=-\,\frac{V_1^2}{4V_2}\left[1-\,\frac{\sqrt{2V_2\,}\alpha}{|V_1|}\,N\right]^2\,.
\end{eqnarray}
with $N=n+\frac{1}{2}\,\,(n=0, 1, 2, \ldots)$. The solutions of the Morse problem is studied in \cite{LE}, and the results are also used in \cite{DRA}.

For presenting the similarities we apply the Langer transformation \cite{NC}
\begin{eqnarray}
\label{eq11}
\psi_{n\ell}=e^{-\Gamma\alpha x}g\,,\,\,e^{-\Gamma\alpha x/2}=\sqrt{\frac{r}{r_0}\,}\,,\,\,(r_0>0,\,\Gamma>0)\,,
\end{eqnarray}
on the equations obtained from the eigenvalue equation, $h_0\psi_{n\ell}=E_{n}^{(0)}\psi_{n\ell}$, which gives
\begin{eqnarray}
\label{eq12}
\left\{-\frac{1}{2}\frac{d^2}{dx^2}+\frac{1}{2}\Gamma^2\alpha^2\left[\nu+(\ell+\frac{1}{2})^2\right]+(\Gamma\alpha r_0)^2\left[sr_0^{\mu}e^{-\Gamma\alpha(\mu+2)x}-E_{n}^{(0)}e^{-2\Gamma\alpha x}\right]\right\}g(x)=0\,.
\end{eqnarray}
The similarity with Eq. (\ref{eq7}) may be give unperturbed eigenvalues for harmonic oscillator- and Coulomb-like potentials by setting $\mu=2,\,\Gamma=\frac{1}{2}$ and $\mu=-1,\,\Gamma=1$, respectively. Details about other combinations and some restrictions on parameter(s) can be found in \cite{NC}.

Writing $\mu=2$ and $\Gamma=\frac{1}{2}$ and comparing Eq. (\ref{eq12}) with Eq. (\ref{eq7}) gives $V_1=-\alpha^2r^2_0E_{n}^{(0)}/4$ and $V_2=-\alpha^2r_0^4s/4$. With the help of Eq. (\ref{eq10}) we obtain unperturbed eigenvalues for harmonic oscillator-type potential as
\begin{eqnarray}
\label{eq13}
E_{n}^{(0)}=\sqrt{2s\,}(2n+1+\sqrt{A\,})\,,\,\,\,A=\nu+(\ell+\frac{1}{2})^2\,,
\end{eqnarray}
and the corresponding unperturbed wave functions as
\begin{eqnarray}
\label{eq14}
\psi_{n\ell}(r)=N_nr^{\frac{1}{2}+\sqrt{A\,}}e^{-\sqrt{\frac{s}{2}\,}r^2}L_n^{(\sqrt{A\,})}(\sqrt{2s\,}r^2\,)\,,
\end{eqnarray}
with normalization constant $N_n$, and in terms of $y=\sqrt{2s\,}r^2$ as
\begin{eqnarray}
\label{eq15}
\psi_{n\ell}(y)=\Lambda(n)y^{\frac{1}{2}(\frac{1}{2}+\sqrt{A\,})}e^{-\frac{y}{2}}L_{n}^{(\sqrt{A\,})}(y)\,.
\end{eqnarray}
with $\Lambda(n)=\sqrt{2(2s)^{1/4}\,}\sqrt{\frac{\Gamma(n+1)}{\Gamma(n+\sqrt{A\,}+1)}\,}$.

Following the same steps we list the solutions for the Coulomb-type potential as
\begin{eqnarray}
\label{eq16}
&&E_{n}^{(0)}=-\left[\frac{\sqrt{2\,}Z}{N+\sqrt{A\,}}\right]^2\,, \\
&&\psi_{n\ell}(r)=N_n r^{\frac{1}{2}+\sqrt{A\,}}e^{-\frac{\Delta_n}{2}r}L_n^{(2\sqrt{A\,})}(\Delta_n r)\,,\,\,\,\Delta_n=\frac{4Z}{2n+1+2\sqrt{A\,}}\,,
\end{eqnarray}
and in terms of a new variable $r=y/\Delta_n$
\begin{eqnarray}
\label{eq17}
\psi_{n\ell}(y)=\Lambda(n)y^{\frac{1}{2}+\sqrt{A\,}}e^{-\frac{y}{2}}L_{n}^{(2\sqrt{A\,})}(y)\,.
\end{eqnarray}
with $\Lambda(n)=\sqrt{\frac{\Gamma(n+1)}{4Z\Gamma(n+2\sqrt{A\,}+1)}\,}$.

According to Eqs. (\ref{eq13}) and (\ref{eq16}) the dependence of unperturbed eigenvalues for harmonic oscillator-like potential on the screening parameter is explicit, as expected, while the unperturbed eigenvalues of Coulomb-like potential are dependent on $Z$, but not $s$. In the next Section, we see that the dependence of the eigenvalues of the harmonic oscillator potential on the parameter $s$ appears when the small corrections are taken into account. We obtain first- and second-order corrections with the help of some equalities and analytic calculations of some integrals of generalized Laguerre polynomials, where we write the initial perturbing term as $h'=\omega^2/r^2$ for harmonic oscillator-type, and as $h'=\omega^2/r$ for Coulomb-type potential, the parameter $\omega^2$ is real, in principle.

\section{Perturbative Solutions}
We use the following definitions to calculate the small corrections \cite{EL, AA}
\begin{subequations}
\begin{align}
\label{eq18}
E_{n}^{(1)}&=<n|(Z-s)r^{\mu}|n>\,,\\
\label{eq19}
E_{n}^{(2)}&=<n|h'|n>+\sum_{m \neq n}\frac{|<m|(Z-s)r^{\mu}|n>|^2}{E_{n}^{(0)}-E_{m}^{(0)}}\,.
\end{align}
\end{subequations}
where $|n>$ represents normalized, unperturbed eigenfunctions $\psi_{n\ell}$.

In order to obtain the diagonal and non-diagonal matrix elements of the Hamiltonians given in Eq. (\ref{eq6}) for the harmonic oscillator-type potential we have the recursion relation for generalized Laguerre polynomials \cite{TD, AA}
\begin{eqnarray}
\label{eq20}
yL_n^{(k)}=(2n+k+1)L_n^{(k)}-(n+1)L_{n+1}^{(k)}-(n+k)L_{n-1}^{(k)}\,,
\end{eqnarray}

Using Eq. (\ref{eq20}) in Eqs. (\ref{eq18}) and (\ref{eq19}), and with the help of Eq. (\ref{eq15}) we write
\begin{eqnarray}
\label{eq21}
y|n>=(2n+1+\sqrt{A\,})|n>-\sqrt{(n+1)(n+1+\sqrt{A\,})\,}|n+1>-\sqrt{n(n+\sqrt{A\,})\,}|n-1>\,,\nonumber\\
\end{eqnarray}
where $|n>$ represents the unperturbed, normalized eigenfunctions of harmonic oscillator-type potential given in Eq. (\ref{eq15}). We obtain the first-order corrections to energy levels by using Eq. (\ref{eq21}) as
\begin{eqnarray}
\label{eq22}
E_{n}^{(1)}=(Z-s)<n|r^2|n>=\frac{Z-s}{\sqrt{2s\,}}<n|y|n>=\frac{Z-s}{\sqrt{2s\,}}(2n+1+\sqrt{A\,})\,.
\end{eqnarray}

To calculate the first term in Eq. (\ref{eq19}) having diagonal elements of the Hamilton operator we have to obtain a type of integrals including the square of generalized Laguerre polynomials. One can define these polynomials in terms of generating function as
\begin{eqnarray}
\label{eq23}
\frac{1}{(1-f)^{\sigma+1}}\,e^{-\,\frac{fz}{1-f}}=\sum_{i=0}^{\infty}L_{i}^{(\sigma)}(z)f^i=U_{i}(z,f)\,,
\end{eqnarray}
where should be $|f|<1$ \cite{TD}. If the left hand side is written as a power series of $f$ the generalized Laguerre polynomials can be given clearly
\begin{eqnarray}
\label{eq24}
L_p^{(q)}(z)=\sum_{j=0}^r\Big(\begin{array}{c}
   p+q \\
  p-j
\end{array}\Big)\frac{(-z)^j}{j!}\,.
\end{eqnarray}
where $\Big(\begin{array}{c}
   a \\
  b
\end{array}\Big)=\frac{a!}{b!(a-b)!}$ are the Binom coefficients \cite{TD}. Writing Eq. (\ref{eq23}) for $L_n^{(k)}(z)$, and $L_m^{(k)}(z)$, separately, multiplying side by side, multiplying both sides of the result by $e^{-z}z^{\sigma}$, and integrating the left and right sides over $z$ for $0<z<\infty$, we obtain \cite{AA}
\begin{eqnarray}
\label{eq25}
\int_{0}^{\infty}z^{k-1}e^{-x}L_{m}^{(k)}(z)L_{n}^{(k)}(z)dz=\left\{
  \begin{array}{ll}
    -\frac{\Gamma(n+k)(n+1)}{\Gamma(n+2)} & {\scriptstyle m=n-1;} \\
    \frac{\Gamma(n+k)}{\Gamma(n+2)} {\scriptstyle (2n+k+1)} & {\scriptstyle m=n;} \\
    -\frac{\Gamma(n+k+1)}{\Gamma(n+2)} & {\scriptstyle m=n+1}.
  \end{array}
\right.
\end{eqnarray}
Inserting the above result into Eq. (\ref{eq19}), the first term is
\begin{eqnarray}
\label{eq26}
<n|\frac{\omega^2}{r^2}|n>=\sqrt{2s\,}\omega^2<n|\frac{1}{y}|n>=\sqrt{2s\,}\omega^2\,\frac{2n+1+\sqrt{A\,}}{(n+\sqrt{A\,})(n+1)}\,,
\end{eqnarray}

We also need the non-diagonal terms of the Hamilton operator in Eq. (\ref{eq19}), which is given as
\begin{eqnarray}
\label{eq27}
\sum_{m \neq n}\frac{|<m|(Z-s)r^{2}|n>|^2}{E_{n}^{(0)}-E_{m}^{(0)}}=(2n+1+\sqrt{A\,})\left[\frac{\sqrt{2s\,}\omega^2}{(n+1)(n+\sqrt{A\,})}-\frac{(Z-s)^2}{4s\sqrt{2s\,}}\right]\,,
\end{eqnarray}

The results given in Eqs. (\ref{eq22}), (\ref{eq26}) and (\ref{eq27}), and with the help of Eq. (\ref{eq13}) makes it possible to write the energy eigenvalues including first- and second-order corrections of the harmonic oscillator-like potential
\begin{eqnarray}
\label{eq28}
E_n=E_n^{(0)}+E_n^{(1)}+E_n^{(2)}&=&\sqrt{2s\,}\left[2n+1+\sqrt{A\,}+\omega^2\frac{2n+1+\sqrt{A\,}}{(n+\sqrt{A\,})(n+1)}\right] \nonumber\\&-&\frac{5s^2+Z^2-6Zs}{4s\sqrt{2s\,}}(2n+1+\sqrt{A\,})\,.
\end{eqnarray}
It seems that the total eigenvalues in Eq. (\ref{eq28}) get gradually decreased for $s \ll 1$ as expected. The other case ($s \gg 1$) is not considered because this means going beyond its physical purpose.

For the case where we take the parameter values of $\mu$ and  $\Gamma$ for the Coulomb-type potential Eq. (\ref{eq18}) gives the first-order correction as
\begin{eqnarray}
\label{eq29}
E_{n}^{(1)}=(Z-s)<n|\frac{1}{r}|n>=(Z-s)\Delta_n<n|\frac{1}{y}|n>=4Z(Z-s)\frac{1}{(2n+1+2\sqrt{A\,})^2}\,,
\end{eqnarray}
where $|n>$ represents the unperturbed, normalized eigenfunctions of Coulomb-type potential given in Eq. (\ref{eq17}), and we need the following integral
\begin{eqnarray}
\label{eq30}
\int_{0}^{\infty}z^{k}e^{-z}L_{m}^{(k)}(z)L_{n}^{(k)}(z) dz=\frac{\Gamma(n+k+1)}{\Gamma(n+1)}\,\delta_{mn}\,,
\end{eqnarray}
which is obtained by following the same steps explained above, and also known as the orthogonality relation of Laguerre polynomials. Eq. (\ref{eq19}) gives us
\begin{eqnarray}
\label{eq32}
<n|\frac{\omega^2}{r}|n>=\omega^2\Delta_n<n|\frac{1}{y}|n>=\frac{4Z\omega^2}{(2n+1+2\sqrt{A\,})^2}\,,
\end{eqnarray}

Finally, we obtain the perturbed energy eigenvalues of the Coulomb-type potential including the diagonal and non-diagonal matrix elements with the help of Eq. (\ref{eq16}) as
\begin{eqnarray}
\label{eq33}
E_n&=&E_n^{(0)}+E_n^{(1)}+E_n^{(2)}\nonumber \\&=&\frac{2(Z-s)^2}{(2n+1+2\sqrt{A\,})^2}+Z\left(\frac{2\omega}{2n+1+2\sqrt{A\,}}\right)^2-2\left(\frac{s}{2n+1+2\sqrt{A\,}}\right)^2\,.
\end{eqnarray}
The limitation on the potential parameter $\nu$ studied in Ref. \cite{NC} remains unchanged, and the control parameter $s$ appears only in one term in Eq. (\ref{eq33}). One expects that the total energy eigenvalue decreases as the parameter's value is reduced.

Eqs. (\ref{eq28}) and (\ref{eq33}) are also valid pure harmonic oscillator potential ($\nu=0$) and pure Coulomb potential ($\nu=0$). So, we are able to give the energy eigenvalues including first- and second-order corrections of these two potentials. For the first potential we obtain
\begin{eqnarray}
\label{eq34}
E_{n}=(n'+\frac{3}{2})\left[\sqrt{2s\,}(1+\frac{2\omega^2}{(n+1)(n'+\ell+1)})-\frac{5s^2+Z^2-6Zs}{4s\sqrt{2s\,}}\right]\,,
\end{eqnarray}
 for the last one
\begin{eqnarray}
\label{eq35}
E_{n}=2Z\left[\frac{\sqrt{Z-2s+2\omega^2\,}}{n'+\ell+2}\right]^2\,.
\end{eqnarray}
with $n'=2n+\ell$.

We compare numerical results with the ones obtained within the Lagrange-mesh formalism. Among the others, this formalism gives numerical results without analytical computations, and has a fast convergence \cite{DB, DB1}. The results presented in Tables ($\Omega$ represents $\omega^2$, here) show that they have an acceptable accuracy. In Tables, "$E_n^{pure}$" represents the results obtained from Eq. (\ref{eq34}) for the harmonic oscillator-type potential, and the ones from Eq. (\ref{eq35}) for the Coulomb-like potential, while "$E_n$" represents the results obtained from Eqs. (\ref{eq28}) and (\ref{eq33}) for the potentials, respectively.

\section{Conclusions}
We present the energy eigenvalues with first- and second-order corrections within perturbation theory for three-dimensional harmonic oscillator- and Coulomb-like potentials, separately. We give some integrals of generalized Laguerre polynomials with analytical result. The unperturbed bound state eigenvalues and the corresponding eigenfunctions of the above potentials are obtained from the ones of one-dimensional generalized Morse potential by the help of some similarities and using the Langer transformation. The effect of screening parameter $s$ on perturbative corrections is discussed.


\newpage

\begin{table}
\begin{ruledtabular}
\caption{Numerical energies of harmonic oscillator-type potential for two different values of parameter $s$.}
\begin{tabular}{@{}ccccccc@{}}
&&&&&{$s=0.250$} &{$s=0.500$} \\ \cline{6-7}
$\nu$ & $\Omega$ & $n$ & $\ell$ & Ref. \cite{DB} & &    \\
0.8 & 0.0001 & 0 & 0 & 1.49999999999999 & $E_{n}^{pure}$\,\, 1.3421223038141774087 & 1.5002999999999999670 \\
 & & & &  & $E_{n}$\,\, 1.8114456625918151644 & 2.0248926666032551758 \\
 & & 1 & 0 & 3.49999999999999 & $E_{n}^{pure}$\,\, 3.1312062299440550639 & 3.5001166666666665428 \\
 & & & & & $E_{n}$\,\,  3.6005897871821628264 & 4.0247944667491521997 \\
 & & 2 & 0 & 5.50000000000000 & $E_{n}^{pure}$\,\, 4.9203891510232988793 & 5.5000733333333329256 \\
 & & & & & $E_{n}$\,\, 5.3897800181172828005 & 6.0247614711131376453 \\
 & 0.01 & 0 & 0 & 1.49999999999999 & $E_{n}^{pure}$\,\, 1.3631233752154177097 &  1.5300000000000000266 \\
 & & & & & $E_{n}$\,\, 1.8252776687630511976 &  2.0444540773254455956 \\
 & & 1 & 0 & 3.49999999999999  & $E_{n}^{pure}$\,\, 3.1393733132667600572 & 3.5116666666666667140 \\
 & & & & & $E_{n}$\,\, 3.6075474528613753478 & 4.0346340919151524318 \\
 & & 2 & 0 & 5.50000000000000 & $E_{n}^{pure}$\,\, 4.9255227462547130912 & 5.5073333333333334139 \\
 & & & & & $E_{n}$\,\, 5.3944278714369371386 & 6.0313345283137138608 \\
 & 0.0001 & 1 & 1 & --- & $E_{n}^{pure}$\,\, 4.0257941549497706646 & 4.5000900000000001455 \\
 & & & &  & $E_{n}$\,\, 4.2462450213022844281 & 4.7465113306289890005 \\
 & & 2 & 1 & --- & $E_{n}^{pure}$\,\, 5.8149878509894898926 & 6.5000619047619050406 \\
 & & & & & $E_{n}$\,\,  6.0354399262919837810 & 6.7464849451043518158 \\
 & & 2 & 2 & --- & $E_{n}^{pure}$\,\, 6.7095901426091728581 & 7.50005555555555559195 \\
 & & & & & $E_{n}$\,\, 6.8484179281510213499 & 7.6552384242335023146 \\
 & 0.01 & 1 & 1 & --- & $E_{n}^{pure}$\,\, 4.0320944763701422886 & 4.5090000000000003411 \\
 & & & & & $E_{n}$\,\, 4.2522940979233672110 & 4.7550660168263583216 \\
 & & 2 & 1 & ---  & $E_{n}^{pure}$\,\, 5.8193214054056179307 & 6.5061904761904765238 \\
 & & & & & $E_{n}$\,\, 6.0396419219568642234 & 6.7524274643626602810 \\
 & & 2 & 2 & --- & $E_{n}^{pure}$\,\, 6.7134792299056984533 & 7.5055555555555555358 \\
 & & & & & $E_{n}$\,\, 6.8522551573336816233 & 7.6606650857855536074 \\
\end{tabular}
\end{ruledtabular}
\end{table}

\newpage

\begin{table}
\begin{ruledtabular}
\caption{Numerical energies of Coulomb-type potential.}
\begin{tabular}{@{}cccccccc@{}}
$\nu$ & $\Omega$ & $s$ & $n$ & $\ell$ & Ref. \cite{DB1} & $E_{n}^{pure}$ &  $E_{n}$  \\
0.1 & 1 & 0.001 & 0 & 0 & -0.4999999999999997 & -0.49899999999999999911 & -0.41876188233291583574 \\
    &   &       & 1 & 0 & -0.1249999999999999 & -0.12474999999999999978 & -0.11406173309417935724 \\
    &   &       & 2 & 0 & -0.05555555555555552 & -0.055444444444444442033 & -0.052207356912285146633 \\
    &   &       & 1 & 1 & -0.05555555555555552 & -0.055444444444444442033 & -0.054245540164301886410 \\
    &   &       & 0 & 2 & -0.05555555555555564 & -0.055444444444444442033 & -0.054715388438544898531 \\
    &   & 0.0001 & 0 & 0 & -0.4999999999999997 & -0.49990000000000001101 & -0.41951716428501928391 \\
    &   &       & 1 & 0 & -0.1249999999999999 & -0.12497500000000000275 & -0.11426745565887827527 \\
    &   &       & 2 & 0 & -0.05555555555555552 & -0.055544444444444444897 & -0.052301518477858410794 \\
    &   &       & 1 & 1 & -0.05555555555555552 & -0.055544444444444444897 & -0.054343377811892815132 \\
    &   &       & 0 & 2 & -0.05555555555555564 & -0.055544444444444444897 & -0.054814073507872942970 \\
\end{tabular}
\end{ruledtabular}
\end{table}

\end{document}